\documentclass[%
 reprint,
superscriptaddress,
 amsmath,amssymb,
 aps,
]{revtex4-1}

\usepackage{graphicx}% Include figure files
\usepackage{dcolumn}% Align table columns on decimal point
\usepackage{bm}% bold math
\usepackage{color}
\usepackage[colorlinks,bookmarks=false,citecolor=darkblue,linkcolor=red,urlcolor=blue]{hyperref}
\definecolor{darkblue}{rgb}{0,0.02,0.45}

\newcommand{\fmo}{Fe$_{2}$Mo$_3$O$_8$}

\newcommand{\mmo}{Mn$_{2}$Mo$_3$O$_8$}
\newcommand{\TN}{\ensuremath{T_{\mathrm{N}}}}
\newcommand{\ddt}{$d$-$d$ transitions}
\newcommand{\spacegroup}{$P6_{3}mc$}
\newcommand{\Eperp}{$E^{\omega}\perp c$}
\newcommand{\Epara}{$E^{\omega}\parallel c$}

\begin{document}

\title{Structure, phonons, and orbital degrees of freedom in Fe$_2$Mo$_3$O$_8$}
\date{\today}

\author{S.~Reschke}
\affiliation{Experimentalphysik V, Center for Electronic
Correlations and Magnetism, Institute for Physics, Augsburg
University, D-86135 Augsburg, Germany}
\author{A.A.~Tsirlin}
\email{altsirlin@gmail.com}
\author{N.~Khan}
\affiliation{Experimentalphysik VI, Center for Electronic
Correlations and Magnetism, Institute for Physics, Augsburg
University, D-86135 Augsburg, Germany}
\author{L.~Prodan}
\author{V.~Tsurkan}
\affiliation{Experimentalphysik V, Center for Electronic
Correlations and Magnetism, Institute for Physics, Augsburg
University, D-86135 Augsburg, Germany} \affiliation{Institute of
Applied Physics, MD-2028~Chi\c{s}in\u{a}u, Republic of Moldova}
\author{I.~K\'{e}zsm\'{a}rki}
\author{J.~Deisenhofer}
\email{joachim.deisenhofer@physik.uni-augsburg.de}
\affiliation{Experimentalphysik V, Center for Electronic
Correlations and Magnetism, Institute for Physics, Augsburg
University, D-86135 Augsburg, Germany}

\date{\today}

\begin{abstract}
We report on the structural and spectroscopic characterization of the multiferroic \fmo{}. Synchrotron x-ray and neutron diffraction, as well as thermal expansion measurements reveal a lattice anomaly at $\TN{}\simeq 60$\,K but do not show any symmetry lowering in the magnetically ordered state. The lattice parameter $c$ exhibits a non-monotonic behavior with a pronounced minimum around 200\,K, which is also reflected in an anomalous behavior of some of the observed infrared-active optical excitations and parallels the onset of short-range magnetic order.
%, likely promoted by the quasi-2D nature of exchange interactions. 
The infrared reflectivity spectra measured between 5 and 300\,K in the frequency range of $100-8000$\,cm$^{-1}$ reveal most of the expected phonon modes in comparison with the eigenfrequencies obtained by density-functional calculations. The $A_1$ phonons show an overall hardening upon cooling, whereas a non-monotonic behavior is observed for some of the $E_1$ modes. These modes also show a strongly increased phonon lifetime below \TN{}, which we associate with the quenched direction of the orbital moment in the magnetically ordered state. A similar increase is observed in the lifetime of the higher-lying $d$-$d$ excitations of the tetrahedral Fe$^{2+}$ site, which become clearly visible below \TN{} only.
\end{abstract}

\maketitle

\section{Introduction}

\begin{figure}[tb]
\includegraphics[width = 0.8\columnwidth]{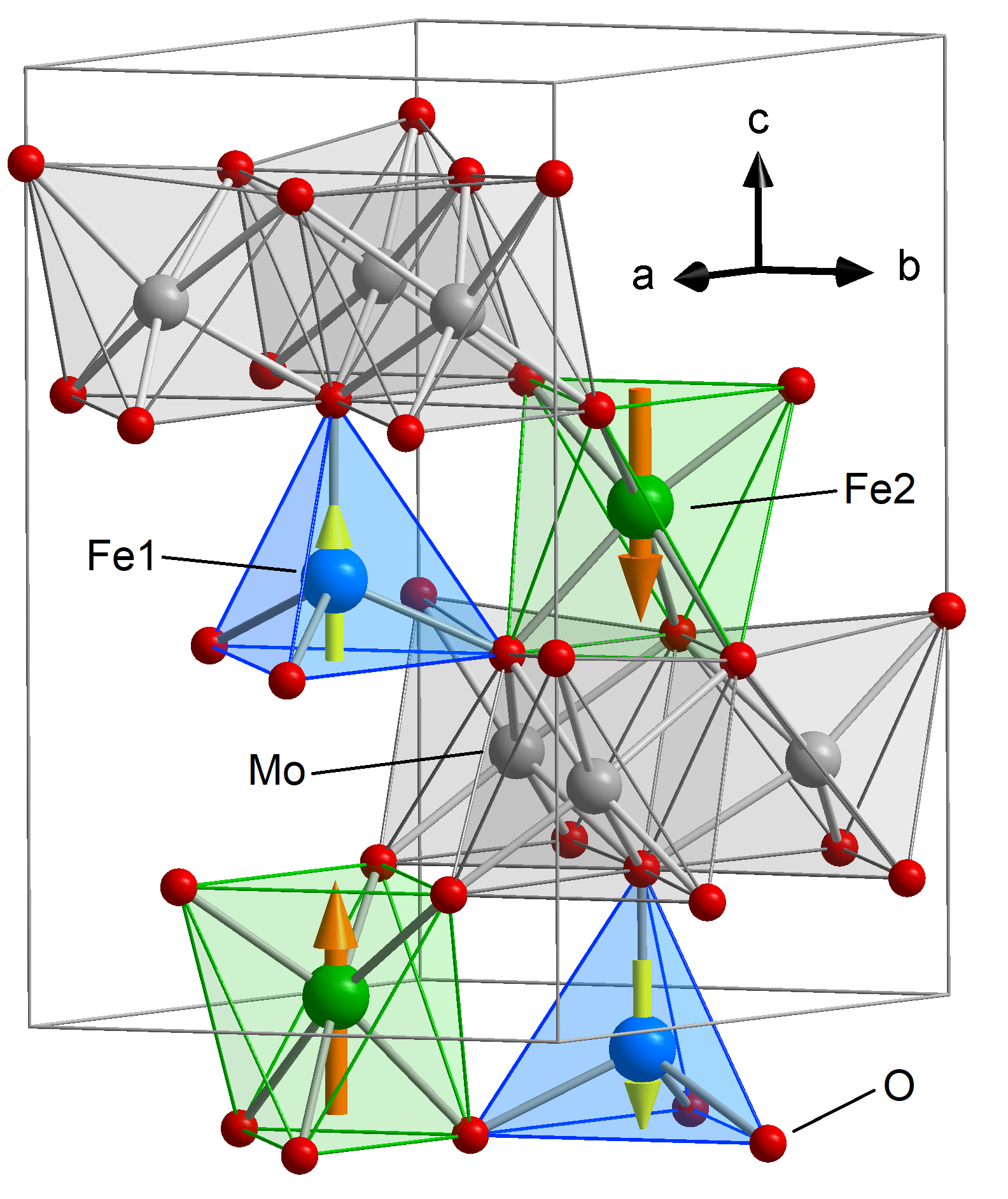}% Here is how to import EPS art
\caption{\label{fig:Crystal_structure} Crystal structure of \fmo{}. Magnetic Fe$^{2+}$ is found in tetrahedral (Fe1) and octahedral (Fe2) oxygen coordination. The arrows indicate the antiferromagnetic spin arrangement.}
\end{figure}

The transition-metal molybdenum oxides $A_{2}$Mo$_3$O$_8$  ($A$ = Mn, Fe, Co) exhibit a hexagonal structure in the polar space group $P6_3mc$ with the resulting polarization pointing along the $c$-axis \cite{McCarroll:1957,Bertrand:1975,LePage:1982,Kurumaji:2015}. The crystal structure of \fmo{} is illustrated in Fig. \ref{fig:Crystal_structure}. The transition-metal ions occupy two different sites with tetrahedral (Fe1) and octahedral (Fe2) coordinations, while the Mo$^{4+}$ ions are located at octahedrally coordinated sites. With the onset of antiferromagnetic ordering of the transition-metal ions, typically occurring between $\TN{}= 40 - 60$\,K \cite{Bertrand:1975}, these compounds become multiferroics. In zero magnetic field, they  realize either a collinear easy-axis antiferromagnetic or a ferrimagnetic state, as observed in \fmo{} \cite{Bertrand:1975,McAlister:1983} and \mmo{} \cite{McAlister:1983,Kurumaji:2017}, respectively. 

The magnetic structure of \fmo{}, with simultaneous antiferromagnetic ordering of the spins on the tetrahedrally and octahedrally coordinated sites, is displayed in Fig. \ref{fig:Crystal_structure}. In the case of \mmo{}, the ferrimagnetic order is realized by ferromagnetic ordering of the spins both at, respectively, tetrahedral and octahedral sites, with the two being antiparallel. The ferrimagnetic and antiferromagnetic states are in fact close in energy. In Fe$_{2}$Mo$_3$O$_8$, the ferrimagnetic spin arrangement can be induced by applying external magnetic field \cite{Wang:2015} or substituting Fe by non-magnetic Zn ions \cite{Kurumaji:2015}. As a clear manifestation of the multiferroic character in the optical properties, magnetoelectric spin excitations were reported in \fmo{} \cite{Kurumaji:2017a} and in Zn-doped \fmo{}, exhibiting an optical diode effect \cite{Yu:2018} and non-reciprocal gyrotropic birefringence \cite{Kurumaji:2017b}, respectively.

Qualitatively, the multiferroic behavior of \fmo{} can be understood within the polar $P6_3mc$ symmetry of its room-temperature crystal structure without invoking symmetry lowering. However, a recent spectroscopic study pinpointed additional IR- and Raman-active, presumably phonon modes appearing below $\TN{}$, and interpreted them as a signature of symmetry lowering \cite{Stanislavchuk:2019}, although no direct measurement of the low-temperature crystal structure was attempted. Moreover, an \textit{ab initio} study~\cite{Solovyev:2019} led to a ground-state spin configuration different from the experimental collinear and commensurate antiferromagnetic order (Fig.~\ref{fig:Crystal_structure}). Taken together, these observations raise the question whether the hexagonal room-temperature crystal structure is the right starting point for modeling the interesting physics of \fmo{}, or effects like charge separation and Jahn-Teller distortions on Fe sites with tetra- and octahedral coordination should be taken into account~\cite{Solovyev:2019}.

In this study, we report on the crystal structure as well as electronic and vibrational IR-active excitations of \fmo{} as a function of temperature. Our diffraction data exclude a symmetry lowering with decreasing temperature, whereas subsequent band-structure calculations suggest that the local trigonal symmetry only partly lifts the orbital degeneracyof the $d^6$ electronic configuration of Fe$^{2+}$. A sizable orbital moment is formed on the tetrahedrally coordinated Fe$^{2+}$ site, whereas vibrational excitations, especially the $E_1$ phonon modes, concurrently show an unusual temperature dependence with a strongly increased lifetime in the magnetically ordered state. Both observations indicate effects beyond spin and lattice degrees of freedom in \fmo{}.

\section{Methods}

Polycrystalline sample of \fmo{} was prepared by a repeated annealing of FeO (99.999\,\%) and MoO$_2$ (99\,\%) in evacuated quartz ampoules at 1000\,$^\circ$C. \fmo{} single crystals were grown by the chemical transport reaction method at temperatures between 950 and 900\,$^\circ$C with TeCl$_4$ as the transport agent. X-ray analysis of the crushed single crystals confirmed single phase of \fmo{}.

X-ray and neutron diffraction experiments were performed on the polycrystalline sample. The neutron data were collected in the temperature range $1.7-275$\,K at the HRPT diffractometer of the Swiss Neutron Source (SINQ) at the Paul Scherrer Institute (Villigen, Switzerland) using neutrons with the 1.494\,\r A wavelength. The sample was placed inside a vanadium can and cooled down in the standard Orange cryostat. High-resolution synchrotron x-ray diffraction was performed at several temperatures between 25\,K and 300\,K at the MSPD beamline~\cite{Fauth:2013} of ALBA (Barcelona, Spain) using the wavelength of 0.3251\,\r A and the multi-analyzer detector. The powder sample was loaded into a spinning glass capillary and cooled down with the He-flow cryostat. The Jana2006 program~\cite{jana2006} was used for the structure refinement.

Thermal expansion measurements were performed with the dilatometer described in Ref.~\onlinecite{Kuechler:2012} using a PPMS from Quantum Design. The length change along the $c$ direction of a hexagonal-shaped crystal was measured. The background contribution has been measured separately and subtracted from the raw data. 

Reflectivity measurements were performed on an as-grown $ab$-plane single crystal and on an $ac$-cut mosaic sample composed of two single crystals. By using the Bruker Fourier-transform IR-spectrometers IFS 113v and IFS 66v/S equipped with He-flow cryostats, the frequency range from 100 to 8000\,cm$^{-1}$ and the temperature range from 5 to 300\,K could be covered. The temperature dependence of eigenfrequencies $\omega_{0}$ and dampings $\gamma$ of the phonon modes was analyzed by an oscillator model with the RefFIT program developed by A. Kuzmenko \cite{Kuzmenko:2018,Kuzmenko:2005}. The optical conductivity was calculated from the reflectivity by Kramers-Kronig transformation with the $\omega^{-1}$ high-frequency extrapolation, followed by the $\omega^{-4}$ extrapolation for frequencies above 800000\,cm$^{-1}$.

Band structures and orbital energies were obtained \textit{ab initio} from relativistic density-functional (DFT) calculations performed in the full-potential FPLO code~\cite{Koepernik:1999}. Additionally, we employed VASP~\cite{Kresse:1996a,Kresse:1996b} for calculating phonon frequencies using frozen displacements. The calculations were performed for the crystal structures refined at 1.7\,K and 275\,K. The Perdew-Burke-Ernzerhof exchange-correlation potential~\cite{Perdew:1996} was combined with the mean-field DFT+$U$ correction for correlation effects in the Fe $3d$ shell. The on-site Coulomb repulsion $U_d=5$\,eV, Hund's coupling $J_d=1$\,eV, and atomic-limit double-counting correction were applied following previous studies~\cite{Xiang:2008}. The calculated phonon frequencies are rather insensitive to the type of magnetic order and to thermal expansion~\cite{supplement}, but drastic changes in the phonon frequencies were observed when DFT was used instead of DFT+$U$, and a non-magnetic (spin-unpolarized) calculation was performed~\cite{supplement}. Such plain-DFT results reproduce the calculated phonon frequencies reported in Ref.~\onlinecite{Stanislavchuk:2019}, but show a quite poor agreement with the experiment~\cite{supplement}.

\section{Results}

\subsection{Crystal structure}
\label{sec:structure}
\begin{figure}
\includegraphics{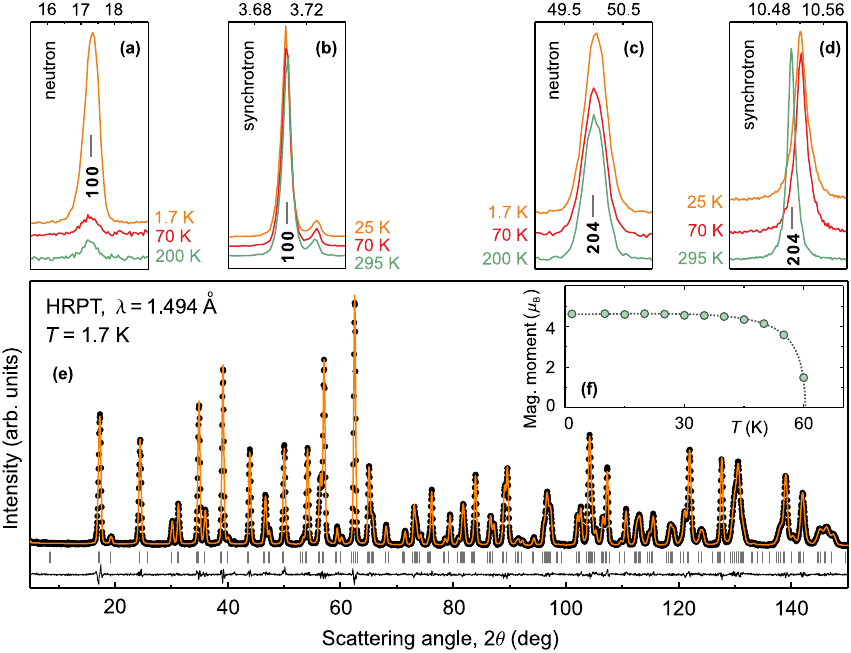}
\caption{(a)--(d) Temperature evolution of the $100$ (a,b) and $204$ (c,d) reflections in the neutron (a,c) and synchrotron (b,d) data. Individual patterns are offset for clarity. (e) Rietveld refinement of the 1.7\,K neutron data using the $P6_3mc$ symmetry of the crystal structure, with the tick marks showing peak positions and the line in the bottom showing the difference pattern. (f) Temperature evolution of the ordered magnetic moment ($\mu_{\text{Fe1}}=\mu_{\text{Fe2}}$) obtained from the neutron data.}
\label{fig:npatterns}
\end{figure}

Both x-ray and neutron diffraction data are compatible with the hexagonal $P6_3mc$ symmetry at all temperatures down to 1.7\,K. Neutron data showed increased intensities of several low-angle reflections, most notably $100$, below $\TN{}$, whereas x-ray intensities remained unchanged (Fig.~\ref{fig:npatterns}). This indicates that the onset of magnetic order in \fmo{} has only a weak influence on the crystal structure, and changes in the neutron data are mostly related to the magnetic scattering. Indeed, neutron diffraction data below $\TN{}$ were successfully refined using the same $P6_3mc$ model (Table~\ref{tab:atomic}) along with the antiferromagnetic structure proposed in Ref.~\onlinecite{Bertrand:1975}, where both Fe1 and Fe2 moments point along the three-fold axis (irreducible representation $\Gamma_5$, magnetic space group $P6_3'm'c$). Adding $ab$-components of the magnetic moment requires admixing the second, irreducible representation $\Gamma_3$, and leads to zero values within the standard deviation. Therefore, we refine the magnetic structure as collinear with the magnetic moments parallel to the $c$-axis. Collinear magnetic structure is further supported by the low-temperature magnetic susceptibility that vanishes for the field applied along the $c$~direction \cite{Strobel:1982}, as expected for a collinear antiferromagnet in a magnetic field applied along the easy axis.

The magnetic structure of \fmo{} entails the Fe1 and Fe2 moments as independent parameters. However, the refinement was not sensitive to their ratio, so we eventually constrained the two and arrived at 4.61(2)\,$\mu_\mathrm{B}$ at 1.7\,K in excellent agreement with the earlier work~\cite{Bertrand:1975}. The ordered moment is somewhat higher than the spin-only value of 4\,$\mu_\mathrm{B}$ for high-spin Fe$^{2+}$, suggesting a sizable orbital contribution.

\begin{figure}
\includegraphics{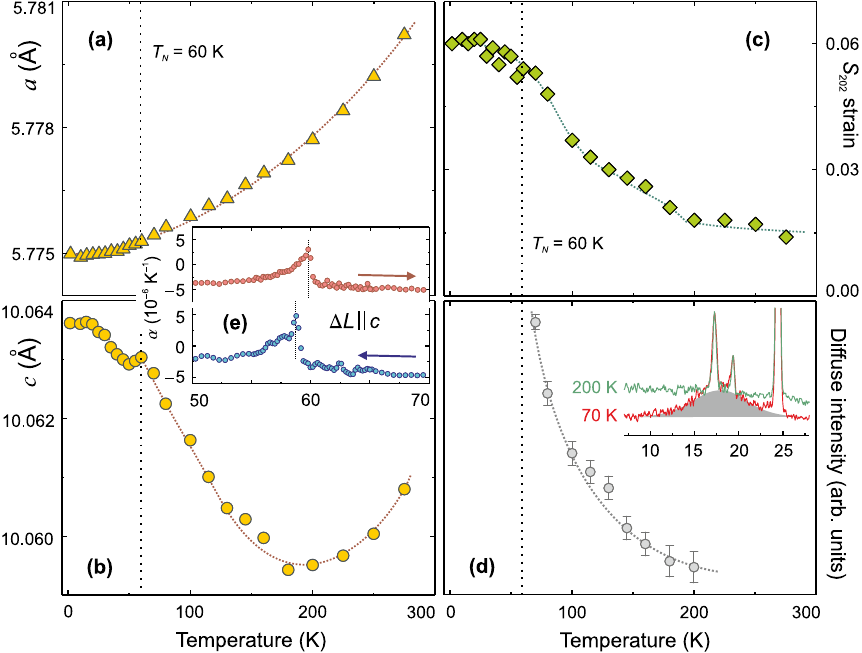}
\caption{(a)--(c) Temperature evolution of the lattice parameters (a,b) and strain parameter $S_{202}$ in $10^{-2}$\,deg$^2$\,\r A$^{-4}$ (c) according to neutron diffraction data. (d) Temperature evolution of the magnetic diffuse scattering. (e) Linear thermal expansion coefficient $\alpha=(1/L_0)(dL/dT)$ determined by dilatometry for $\Delta L\|c$.
}
\label{fig:nparam}
\end{figure}

Neither atomic coordinates nor thermal displacement parameters showed any abrupt changes around $\TN{}$. The atomic coordinates are nearly constant within the temperature range of our study, whereas the displacement parameters systematically increase upon heating~(Table~\ref{tab:atomic}). The in-plane lattice parameter $a$ also increases upon heating due to thermal expansion (Fig.~\ref{fig:nparam}a), but the out-of-plane lattice parameter $c$ reveals a rather unusual behavior with a minimum around 200\,K and a kink at $\TN{}$ (Fig.~\ref{fig:nparam}b). This kink is even better visible in the linear thermal-expansion coefficient obtained from dilatometry (Fig.~\ref{fig:nparam}e). 

\begin{table}
\caption{\label{tab:atomic}
Structural parameters of Fe$_2$Mo$_3$O$_8$ at 1.7\,K (upper rows) and 275\,K (bottom rows), as determined from Rietveld refinements of the neutron diffraction data. The atomic displacement parameters $U_{\rm iso}$ are given in $10^{-2}$\,\r A$^2$. The lattice parameters are $a=5.77499(2)$\,\r A, $c=10.0636(1)$\,\r A at 1.7\,K and $a=5.78021(4)$\,\r A, $c=10.0608(2)$\,\r A at 275\,K, and the space group is $P6_3mc$. The refinement residuals $R_I/R_p$ are $0.009/0.034$ at 1.7\,K and $0.011/0.056$ at 275\,K, with the higher $R_p$ due to the lower statistics of the 275\,K data.
}
\begin{ruledtabular}
\begin{tabular}{cccccc}
    & site & $x/a$ & $y/b$ & $z/c$ & $U_{\rm iso}$ \\\hline
Fe1 & $2b$ & $\frac13$ & $\frac23$ & 0.95310(1)    & 0.12(2) \\
    &      &           &           & 0.95315(1)    & 0.50(4) \\
Fe2 & $2b$ & $\frac13$ & $\frac23$ & 0.51260(1)    & 0.17(2) \\
    &      &           &           & 0.51257(1)    & 0.63(4) \\
Mo  & $6c$ & 0.1461(1) & $-x$    & 0.25\footnotemark & 0.02(2) \\
    &      & 0.1463(2) &           & 0.25\footnote[1]{Fixed to define unit cell origin}  & 0.25(3) \\
O1  & $2a$ & 0         & 0         & 0.39048(1)    & 0.29(4) \\
    &      &           &           & 0.39051(1)    & 0.51(7) \\
O2  & $2b$ & $\frac13$ & $\frac23$ & 0.14643(1)    & 0.46(4) \\
    &      &           &           & 0.14649(1)    & 0.61(7) \\
O3  & $6c$ & 0.4880(1) & $-x$      & 0.36299(2)    & 0.29(2) \\
    &      & 0.4874(2) &           & 0.36288(1)    & 0.59(4) \\
O4  & $6c$ & 0.1665(1) & $-x$      & 0.63421(1)    & 0.31(2) \\
    &      & 0.1664(2) &           & 0.63409(1)    & 0.61(4) \\
\end{tabular}
\end{ruledtabular}
\end{table}
A weak lattice anomaly at $\TN{}$ is not unexpected, but the minimum in the $c$ parameter at 200\,K has no counterpart in thermodynamic or spectroscopic measurements reported so far. A closer inspection of the data suggested two further effects that appear below this temperature. First, reflection broadening becomes more anisotropic, as can be seen from the increase in the $S_{202}$ parameter (Fig.~\ref{fig:nparam}c) which accounts for the anisotropic contribution to the reflection width due to strain~\cite{Stephens:1999}. This parameter is non-zero even at room temperature but remains nearly temperature-independent down to 200\,K, increases below this temperature, and saturates below $\TN{}$. The change in the peak width is hardly noticeable in the neutron data but can be recognized in the high-resolution synchrotron data that show slightly broader $h0l$ reflections below 200\,K (Fig.~\ref{fig:npatterns}d).

Second, neutron diffraction data collected right above $\TN{}$ show a broad diffuse feature around the position of the $100$ reflection. This reflection then attains the largest magnetic contribution in the ordered state, so the broad feature represents magnetic diffuse scattering. It is gradually suppressed upon heating and fully disappears around $200$\,K, where the background of the neutron diffraction pattern develops a weak downward curvature typical of paramagnetic scattering (Fig.~\ref{fig:nparam}d). 

These observations suggest that short-range magnetic order appears in \fmo{} below 200\,K. Its formation may be linked to the minimum in the $c$ parameter, because the onset of spin-spin correlations will often facilitate lattice expansion if it leads to an increase in the exchange energy and stabilization of magnetic order~\cite{Chatterji:2011}. The increase in the strain broadening $S_{202}$ can have a similar origin~\cite{Senn:2013} and does not indicate macroscopic symmetry lowering. Indeed, in systems with the symmetry lowering an anisotropic strain broadening will usually precede the symmetry-lowering phase transition and diverge upon approaching the transition from above~\cite{Wang:2018}. This does not happen in \fmo{}, where the high-resolution synchrotron data collected both right above and well below $\TN{}$ reveal only a weak broadening of the $h0l$ reflections (Fig.~\ref{fig:npatterns}d). 

Regarding the transition at $\TN{}$, it does not lead to any symmetry lowering either. Dilatometry reveals a weak thermal hysteresis around 60\,K (Fig.~\ref{fig:nparam}e), whereas the linear thermal expansion coefficient develops a $\lambda$-type anomaly typical of a second-order phase transition. Therefore, we interpret the transition at $\TN{}$ as weakly first-order and exclude any significant structural changes upon the formation of the magnetically ordered state in \fmo{}. The absence of symmetry lowering is also consistent with our phonon calculations that yield real frequencies for all 50 $\Gamma$-point phonons expected in the hexagonal structure of \fmo{}~\cite{supplement}.

\subsection{Orbital degrees of freedom}
\label{sec:orbital}
We now use our crystallographic data to assess the orbital degrees of freedom in \fmo{} that features octahedrally and tetrahedrally coordinated Fe$^{2+}$ with the $d^6$ electronic configuration. Both ions should be in a high-spin state according to the large magnetic moment they exhibit, and consequently possess orbital degrees of freedom in the absence of local distortions. These orbital degrees of freedom may or may not be quenched in the real structure owing to weak distortions present therein.

\begin{figure*}
\includegraphics{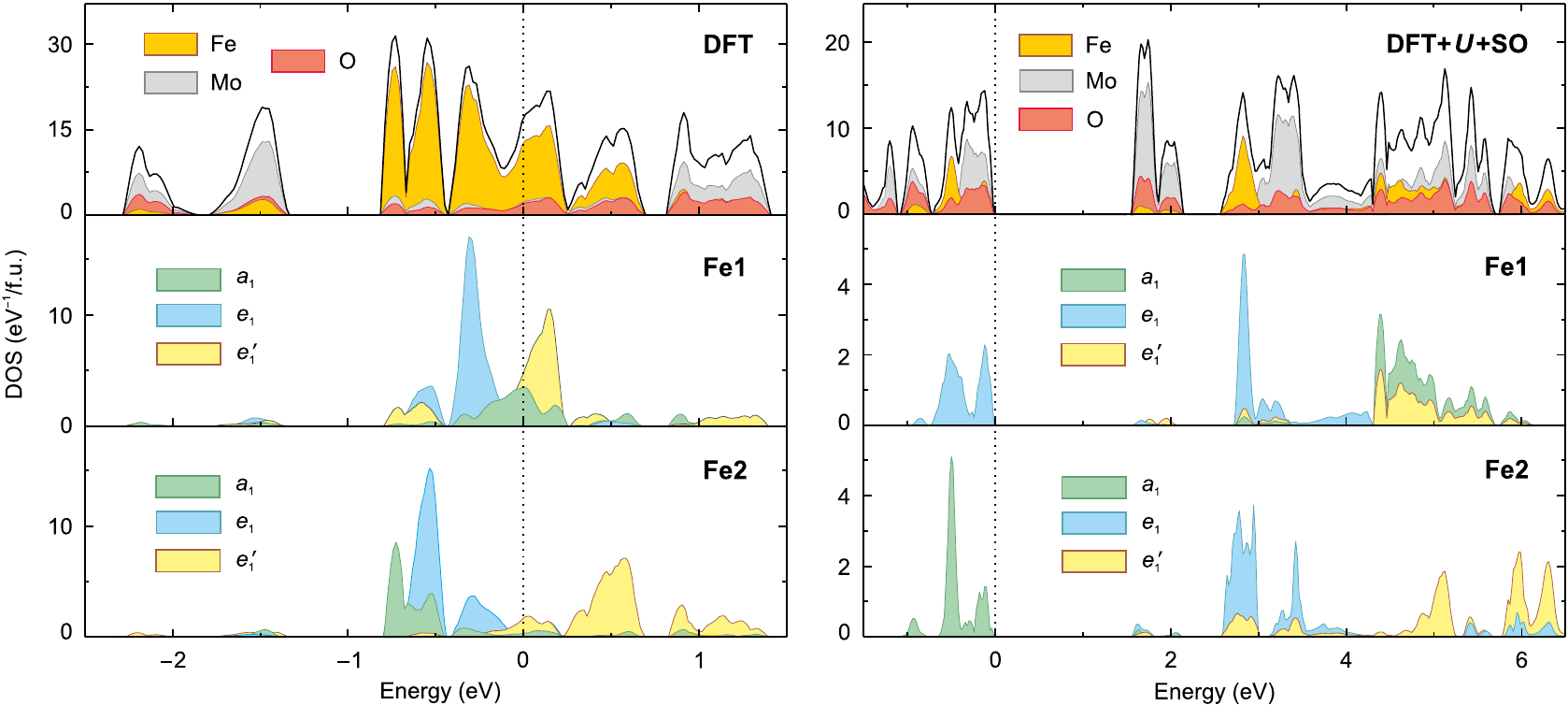}
\caption{\label{fig:dos}
Electronic density of states (DOS) for Fe$_2$Mo$_3$O$_8$: uncorrelated non-magnetic calculation (left) and a DFT+$U$+SO calculation for the antiferromagnetic state with spins along $c$ (right). The upper panels display total and atomic-resolved DOS. The middle and bottom panels display orbital-resolved DOS for Fe1 and Fe2, respectively, with only spin-minority channel shown in the DFT+$U$+SO case. The Fermi level is at zero energy.}
\end{figure*}

The energies of the crystal-field levels in \fmo{} are obtained from DFT by projecting the calculated band structure (FPLO) in the region of $3d$ bands, between $-1$ and $+0.8$\,eV (Fig.~\ref{fig:dos}, left), onto Wannier functions constructed from individual Fe $3d$ orbitals~\cite{Eschrig:2009}, or by calculating centers of gravity for the orbital-resolved density of states (DOS) shown in the left part of Fig.~\ref{fig:dos}. Leading crystal-field splittings are $\Delta_o=0.68$\,eV (0.82\,eV) on the octahedral site and $\Delta_t=0.33$\,eV (0.28\,eV) on the tetrahedral site, where the values in brackets are obtained from the centers of gravity. These values are somewhat lower than $\Delta_o=1.13$\,eV and $\Delta_t=0.50$\,eV reported in Ref.~\onlinecite{Solovyev:2019}. We repeated calculations for the crystal structure used in that work, but obtained the same values as for the 1.7\,K crystal structure from Table~\ref{tab:atomic}. Therefore, we believe that the differences in the orbital energies are caused by the choices of the projection procedure and energy window for the Wannier functions. For example, including the Mo $4d$ band between 0.8 and 1.5\,eV into the Wannier projections will increase $\Delta_o$ and $\Delta_t$.

\begin{figure}
    \centering
    \includegraphics{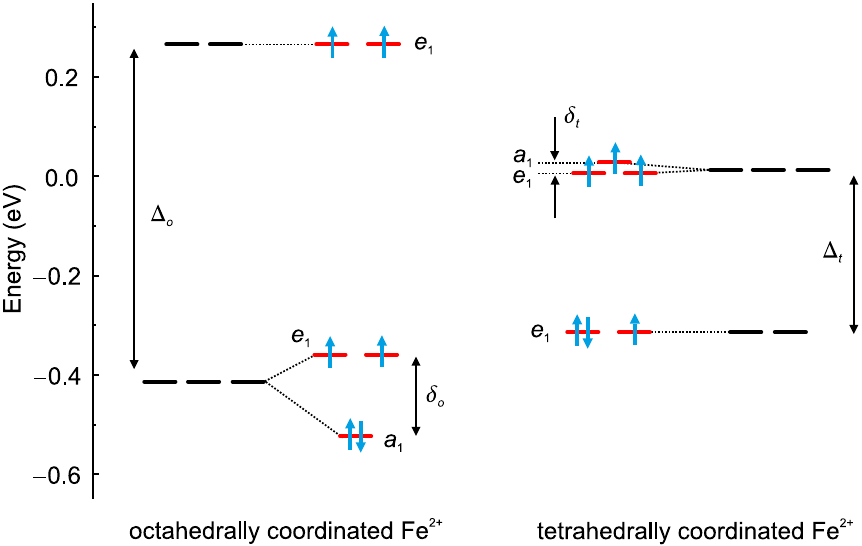}
    \caption{Crystal-field levels of the Fe$^{2+}$ ions in \fmo{}, as determined by DFT using Wannier projections. Orbital energies are given with respect to the Fermi level of the uncorrelated band structure.}
    \label{fig:orbitals}
\end{figure}

The trigonal symmetry of the structure allows a weak distortion and introduces the secondary splittings $\delta_o=0.17$\,eV (0.06\,eV) and $\delta_t=0.02$\,eV (0.01\,eV), where the $a_1$ level is systematically lower in energy for the octahedron and higher in energy for the tetrahedron, as shown in Fig.~\ref{fig:orbitals}. The resulting crystal-field scheme suggests that orbital degeneracy may be lifted for Fe2 if the minority-spin electron occupies the $a_1$ orbital. On the other hand, the lowest level of Fe1 remains doubly-degenerate. Indeed, adding Hubbard $U$ without the spin-orbit (SO) coupling does not lead to a gap opening for the Fe1 states. The band gap of about 1.6\,eV opens only within DFT+$U$+SO, where we obtain an insulating solution shown in the right part of Fig.~\ref{fig:dos}. The minority-spin electron of Fe2 occupies the $a_1$ level, as expected from the crystal-field scheme in Fig.~\ref{fig:orbitals}. In the case of Fe1, the $e_1$ states split into two parts, one above and one below the Fermi level. This splitting caused by the spin-orbit coupling is accompanied by a sizable orbital moment.

We calculate orbital moments for different directions of the spin. The orbital moment of Fe2 is systematically below 0.01\,$\mu_B$, as expected in the absence of orbital degeneracy, whereas the orbital moment of Fe1 equals 0.50\,$\mu_\mathrm{B}$ for spins directed along the $c$-axis and 0.11\,$\mu_B$ for spins lying the $ab$ plane, with the energy gain of 9.7\,meV/f.u. in the former case. These results indicate a large easy-axis anisotropy in agreement with the experimental magnetic structure (Sec.~\ref{sec:structure}) and the highly anisotropic magnetic susceptibility of \fmo{}~\cite{Strobel:1982}. Both spin and orbital moments should preferentially point along the $c$ direction, whereas structural symmetry excludes any canting away from this direction. Indeed, a non-collinear calculation in VASP for the antiferromagnetic state shown in Fig.~\ref{fig:Crystal_structure} returns zero spin components in the $ab$-plane and confirms the collinear nature of the magnetic order in \fmo{}.

\subsection{Optical spectroscopy}
\begin{figure}[tb]
\includegraphics[width = 1\columnwidth]{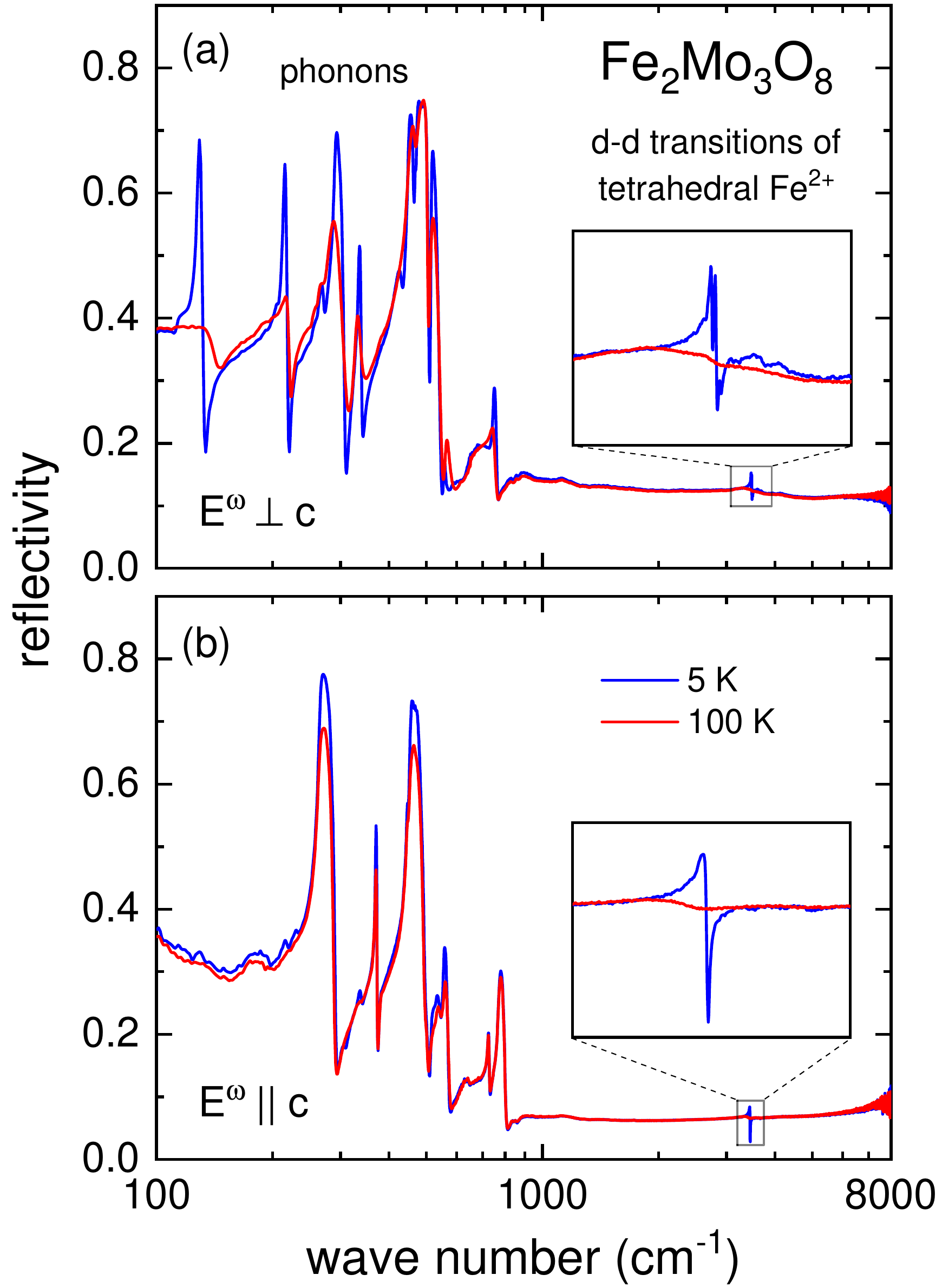}% Here is how to import EPS art
\caption{\label{fig:broadband_R} Reflectivity of \fmo{} for \Eperp{} (a) and \Epara{} (b) at 5\,K and at 100\,K. Below 800\,cm$^{-1}$, the reflectivity is dominated by the phonon modes. The \ddt{} of the tetrahedral Fe$^{2+}$ are located around 3500\,cm$^{-1}$. The insets give a zoom-in of the \ddt{}.}
\end{figure}

Reflectivity spectra of \fmo{} have been measured in the frequency range between 100 and 8000\,cm$^{-1}$ for two polarizations of the incoming light beam, $E^{\omega}\perp c$ and $E^{\omega}\parallel c$. Figure \ref{fig:broadband_R} shows the reflectivity for both polarizations at 5 and 100\,K, i.e. for temperatures below and above the magnetic ordering temperature of $T_{\mathrm{N}}=60$\,K. At frequencies below 900\,cm$^{-1}$, several excitation features including the phonon modes can be observed. Their temperature dependence will be analyzed in detail below. For frequencies around 3500\,cm$^{-1}$ several excitations can be found that broaden considerably above \TN{}. These excitations can be ascribed to \ddt{} of the tetrahedral Fe$^{2+}$ sites, which are commonly found in this frequency range \cite{Low:1960a,Feiner:1982,Mualin:2002,Ohgushi:2005,Fedorov:2006,Laurita:2015,Evans:2017}, whereas the \ddt{} of octahedral Fe$^{2+}$ are usually located at higher energies \cite{Stoehr:2006,Riedel:2007}.

\subsection{\label{sec:phonons}Far-infrared excitations and phonons}

The number and symmetry of the allowed zone-center IR active phonon modes for \fmo{} with space group \spacegroup{} \cite{Bertrand:1975} are given by the irreducible representations of the normal modes, of which the nine one-dimensional $A_1$-modes should be observable for \Epara{}  and the 12 doubly degenerate $E_1$-modes  for \Eperp{}:
\begin{align}
\Gamma &= 9A_1   + 12E_1 &&\text{(Raman- and IR active)}\nonumber \\ 
&+  13E_2 &&\text{(Raman-active)}\nonumber \\  
&+  A_1 +E_1 &&\text{(acoustic)}\nonumber \\ 
&+ 3A_2 + 10B_1 +3B_2  &&\text{(silent).} \nonumber    
\end{align}

\begin{figure}[tb]
\includegraphics[width = 0.981\columnwidth]{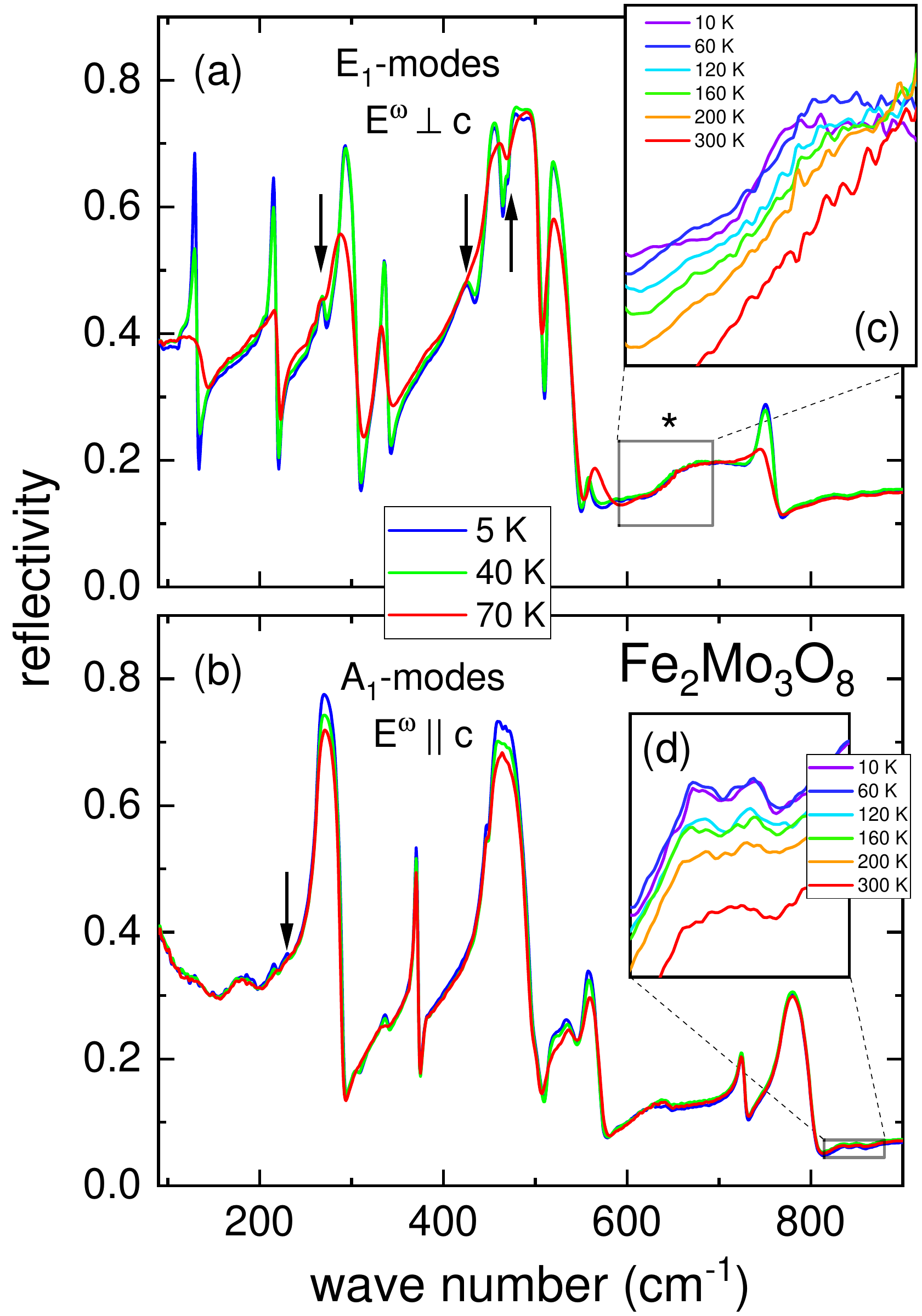}% Here is how to import EPS art
\caption{\label{fig:FIR_R} FIR reflectivity of \fmo{} for (a) polarization \Eperp{}  and (b) \Epara{} at selected temperatures. The arrows indicate additional modes appearing below \TN{}. (c) Temperature dependence of the absorption feature between 600 and 700\,cm$^{-1}$($\ast{}$). (d) Temperature dependence of the weak excitations at approximately 850\,cm$^{-1}$ .}
\end{figure}

\begin{figure}[tb]
\includegraphics[width = 1\columnwidth]{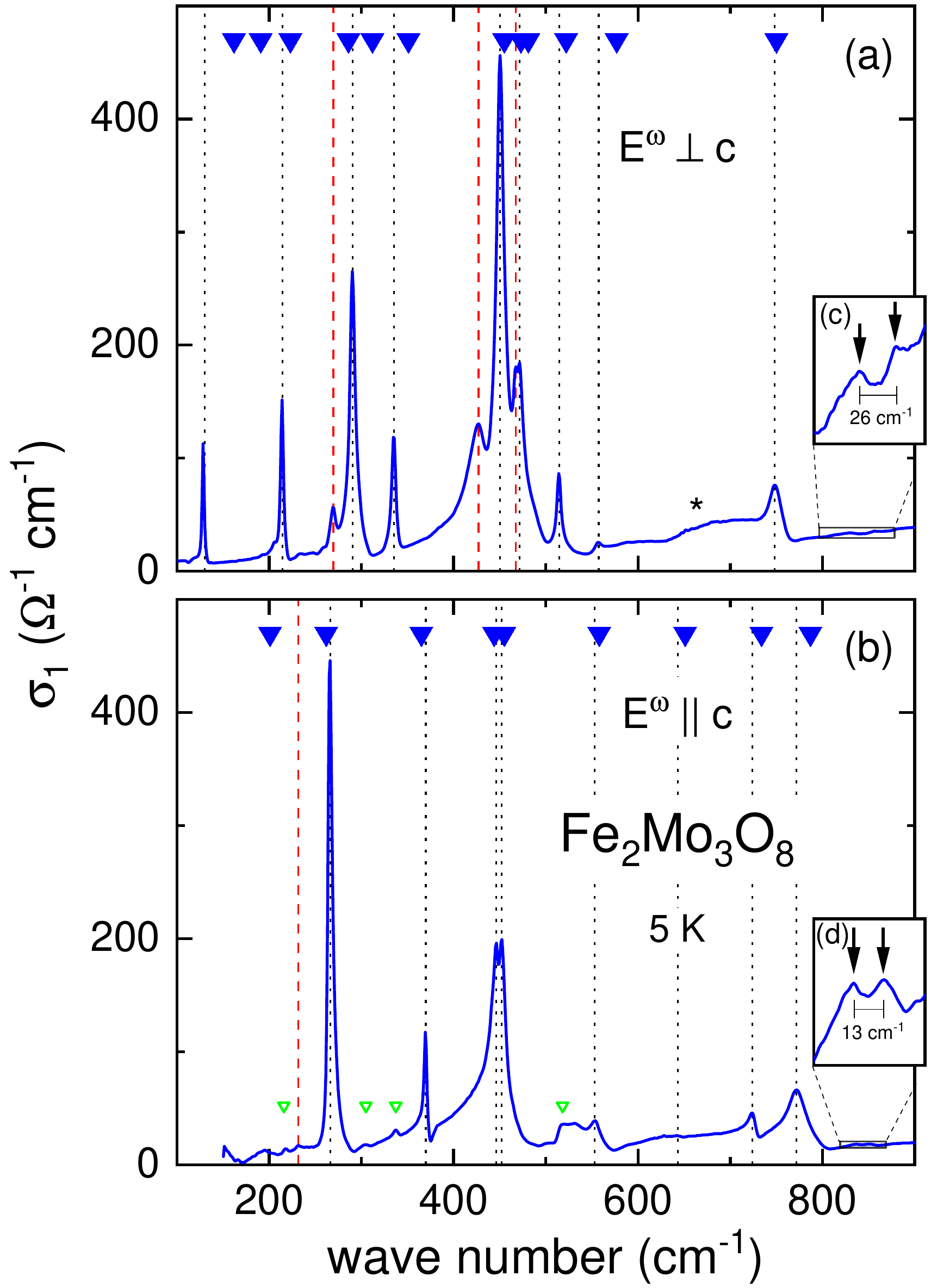}% Here is how to import EPS art
\caption{\label{fig:FIR_sigma} Frequency-dependent optical conductivity at 5\,K for (a) polarization \Eperp{} and (b) \Epara{} (b). Black dotted lines mark the phonons observed for $T>\TN{}$. Additional modes appearing below \TN{} are indicated by red dashed lines. The blue triangles mark the phonon frequencies obtained from DFT+$U$+SO calculations. Green open triangles indicate polarization leakage from \Eperp{}. The asterisk $\ast{}$ indicates the broad feature emerging below 200~K as described in the text. The black arrows in the insets indicate the peaks  (c) at 830 and 856\,cm$^{-1}$ for \Eperp{} and (d) at 837 and 850\,cm$^{-1}$ for \Epara{}.} 
\end{figure}

Figure~\ref{fig:FIR_R} shows the FIR reflectivity spectra of \fmo{} for both \Eperp{} and \Epara{} at three selected temperatures. The optical conductivity calculated via Kramers-Kronig transformation from the 5\,K reflectivity data is shown in Fig.~\ref{fig:FIR_sigma}. For \Epara{}, the reflectivity data above 150\,cm$^{-1}$ has been used for the calculation. 

In the following, we will distinguish three different types of excitation features in the FIR spectra, depending on the temperature range of their occurrence:\\
(i) excitations visible already at room temperature, which are directly compared to IR active phonons of $A_1$ or $E_1$ symmetry, (ii) the spectral feature in the 600-700\,cm$^{-1}$ range shown in detail in Fig.~\ref{fig:FIR_R}(c); this feature occurs below 200\,K concomitantly with the minimum in the $c$-axis lattice parameter (see Fig.~\ref{fig:nparam}), and (iii) excitations observed only in the magnetically ordered state below \TN{} and indicated by the black arrows in Fig.~\ref{fig:FIR_R}.\\
All experimentally observed modes and the calculated phonon frequencies are summarized in Table~\ref{tab:TAB_01}.

We start with the excitations of the first group and compare their frequencies with phonon eigenfrequencies obtained from DFT+$U$+SO calculations. To this end, peaks in the experimental optical conductivity are used, as indicated in Fig.~\ref{fig:FIR_sigma}. Note that the features showing up at 215, 305, 335, and 520\,cm$^{-1}$  for \Epara{ }(indicated by the green arrows in Fig.~\ref{fig:FIR_sigma}) stem from a slight polarization leakage of the \Eperp{} direction. 

For \Eperp{} nine modes are observable below 800\,cm$^{-1}$ already at room temperature and can be compared to the twelve calculated eigenfrequencies of the expected $E_1$ modes in the upper part of Table~\ref{tab:TAB_01}. Similarly, for \Epara{} all observed eight modes present already at room temperature below 800\,cm$^{-1}$ are assigned to the expected $A_1$ phonon modes. In the latter case, there is an excellent agreement between the experimental and calculated frequencies. Only for the lowest-lying $A_1$-mode, we cannot identify an experimental counterpart, because at around 190\,cm$^{-1}$ there is an artefact resulting from our experimental setup. However, no mode was reported in this frequency range in Ref.~\cite{Stanislavchuk:2019}. 

For \Eperp{}, the agreement for the $E_1$ modes is still good but less favorable, and the counterparts of three modes could not be identified experimentally. The excitation features present around 850\,cm$^{-1}$ should not be considered as phonons, even if they are present already at room temperature for both polarizations and have been assigned to a one-dimensional representation $A_1$ in Ref.\;\cite{Stanislavchuk:2019}. First, these features develop a two-peak structure below 200\,K (see Fig.~\ref{fig:FIR_R}(d)), where short-range magnetic order presumably sets in. Second, at the lowest temperatures the energy differences between the two modes (see Figs.~\ref{fig:FIR_sigma}(c) and (d) and Tab.~\ref{tab:TAB_01}) are 13\,cm$^{-1}$ and 26\,cm$^{-1}$ for \Eperp{} and \Epara{}, respectively. These splittings are of the same size as the ones observed for the electronic $d$-$d$ excitations of the tetrahedral Fe$^{2+}$ sites, which will be discussed below. This suggests that the excitations around 850\,cm$^{-1}$ are of electronic origin. Besides, no phonon eigenfrequencies were found in our calculations in this frequency range.

\begin{table}[tb]
\caption{\label{tab:TAB_01}
Summary of experimental excitation frequencies (in cm$^{-1}$) measured at 70\,K and 5\,K. Upper part: Modes already visible at room temperature and corresponding values obtained from the DFT+$U$+SO calculations (1.7\,K crystal structure, antiferromagnetic order). Middle part: excitations concomitant with the onset of short-range magnetic order. Lower part: excitations concomitant with the onset of long-range magnetic order and appearing only below \TN{}.}
\begin{ruledtabular}
\begin{tabular}{cccccc}
\multicolumn{2}{c}{\Eperp{}} & $E_{1}$ & \multicolumn{2}{c}{\Epara{}} & $A_{1}$\\
70\,K& 5\,K & DFT+$U$ & 70\,K & 5\,K & DFT+$U$ \\

\hline
135 &   129 &   162 &       &       &   201 \\
    &       &   191 &   269 &   269 &   262 \\
218 &   214 &   223 &   371 &   371 &   365 \\
    &       &   286 &   447 &   446 &   444 \\
290 &   292 &   312 &   458 &   457 &   454 \\
333 &   335 &   351 &   558 &   556 &   558 \\
454 &   452 &   455 &   643 &   643 &   651 \\
    &       &   473 &   727 &   727 &   734 \\
473 &   471 &   481 &   782 &   782 &   787 \\
510 &   514 &   522 &       &       &       \\
561 &   559 &   577 &       &       &       \\
751 &   754 &   750 &       &       &       \\
    &       &       &        &      &       \\
836 &   837 &       &   831 &  830  &       \\
850 &   850 &       &   857 &  856  &       \\
\hline
613 &   596 &       &       &       &       \\
697 &   692 &       &       &       &       \\
\hline
    &   270 &       &       &   230 &       \\
    &   426 &       &       &       &       \\
    &   468 &       &       &       &       \\
\end{tabular}
\end{ruledtabular}
\end{table}

\begin{figure}[tb]
\includegraphics[width = 1\columnwidth]{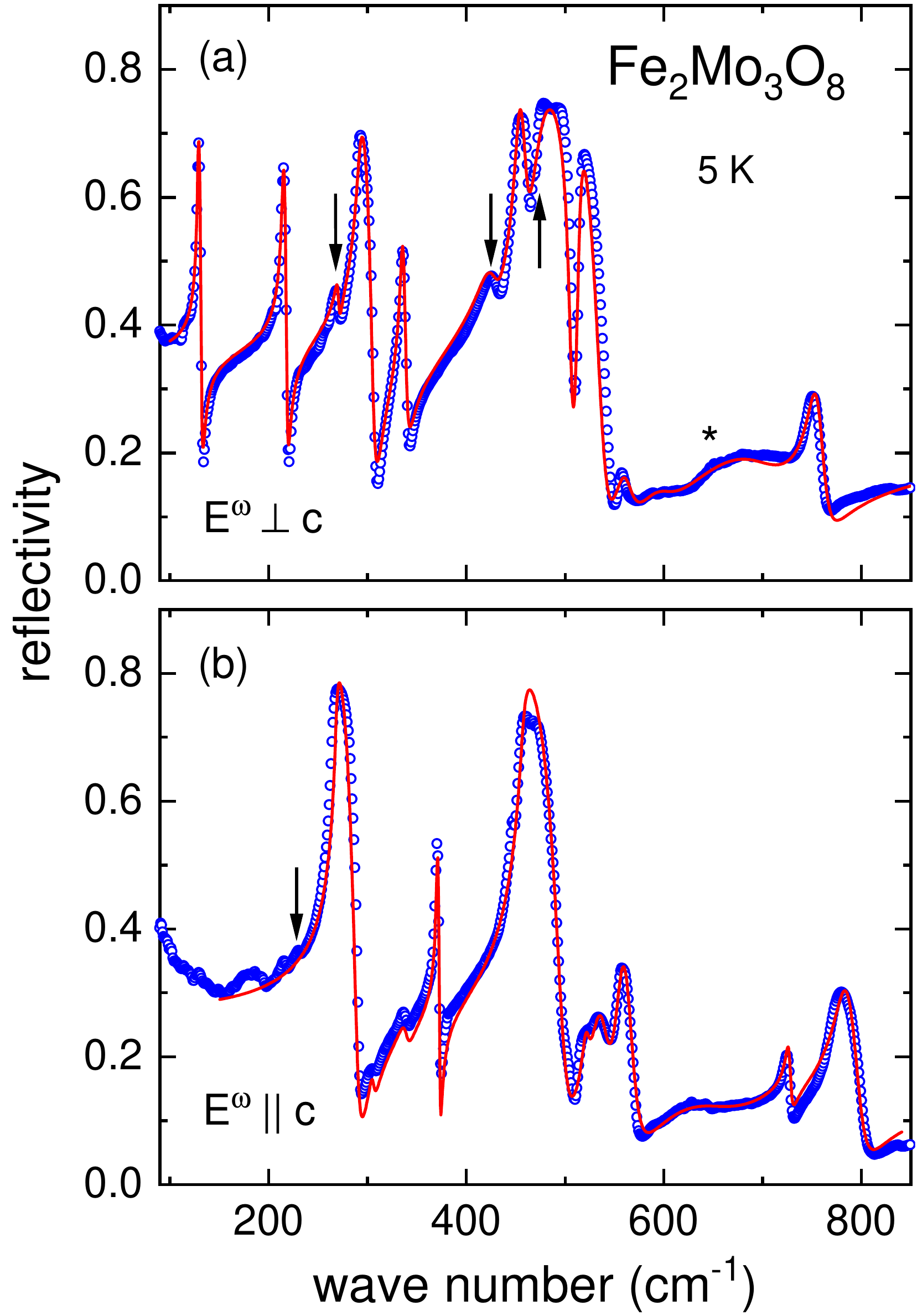}% Here is how to import EPS art
\caption{\label{fig:Fit_examples} Experimental reflectivity spectra measured at 5\,K (open symbols) for \Eperp{} (a) and \Epara{} (b) and corresponding fits (red solid lines) according to the oscillator model described in the text. Arrows indicate the additional modes appearing below \TN{}.}
\end{figure}

The temperature dependence of the modes below 800\,cm$^{-1}$ was obtained by fitting the reflectivity spectra with a sum of Lorentz and Fano oscillators for the complex dielectric constant,
\begin{align}
\epsilon &=\epsilon_\infty \nonumber\\ &+\sum_k\frac{\omega_{p,k}^2}{\omega_{0,k}^2-\omega^2-i\gamma_k \omega}\left( 1+i\frac{\omega_{q,k}}{\omega} \right)^2 + \left( \frac{\omega_{p,k}\omega_{q,k}}{\omega_{0,k}\omega}  \right)^2 \nonumber
\end{align}
Here, $\omega_0$ denotes eigenfrequency, $\gamma$ the damping, and $\omega_p$  the plasma frequency. $\omega_q$ accounts for the asymmetry of the Fano oscillator, which was used to describe the asymmetric lineshapes of the four lowest-lying modes for \Eperp{} and for the 269 and 458\,cm$^{-1}$ modes for \Epara{}. For $\omega_q =0$ the symmetric Lorentzian lineshape is recovered. The Fano lineshape has been utilized to achieve a better fit for these six modes, but it is not clear at present, whether the asymmetry originates from the coupling to continuum background of other degrees of freedom or not.  The values for the high-frequency dielectric constant $\epsilon_\infty$ obtained at the lowest temperature are $7.6$ for \Eperp{} and $5.8$  for \Epara{}.

For \Eperp{} at the highest temperatures the reflectivity was fitted with nine oscillators in agreement with the number of modes discernible at room temperature. Below 200\,K, two additional oscillators were included to take into account the broad features in the region between 600 and 700\,cm$^{-1}$. To describe the modes emerging below \TN{}, two additional modes were included. However, the mode at 468\,cm$^{-1}$ is too weak to be included in the fit (see Fig.~\ref{fig:Fit_examples}). Similarly, for \Epara{} the modes listed in Table~\ref{tab:TAB_01} were included in the fitting procedure, except for the ones at 230 and 447\,cm$^{-1}$, which have an extremely small spectral weight and are close to the experimental resolution limit. Finally,  additional oscillators were included to account for the polarization leakage features.

Figure~\ref{fig:Fit_examples} shows representative fits of the experimental reflectivity data measured at 5\,K for \Eperp{} and \Epara{} with the oscillator model described above. For both polarization directions  the fit describes the experimental data quite well.

\begin{figure}[h!tb]
\includegraphics[width = 1\columnwidth]{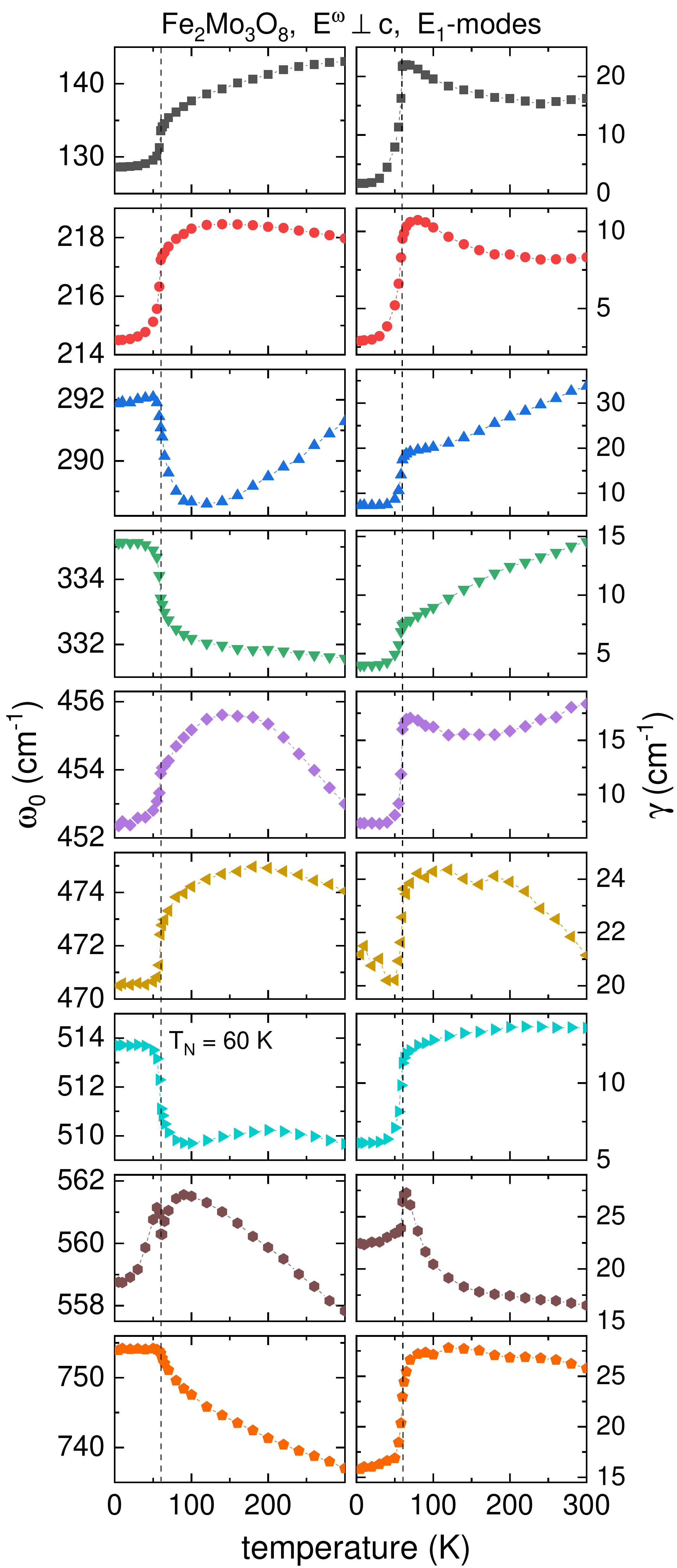}% Here is how to import EPS art
\caption{\label{fig:Fitparameter_Eperp} Temperature dependence of phonon eigenfrequency $\omega_{0}$ and damping $\gamma$ for \Eperp{}. The dashed vertical lines indicate the phase transition at $T_{\mathrm{N}}=60$\,K.}
\end{figure}

The temperature dependence of $\omega_0$ and $\gamma$ of the strongest phonon modes for \Eperp{} is shown in Figure~\ref{fig:Fitparameter_Eperp}. For all these modes clear anomalies of both $\omega_0$ and $\gamma$ can be found at \TN{}. Already above \TN{}, the eigenfrequencies of some modes show a non-monotonic evolution in the temperature range between 150 and 200\,K, correlating with the minimum observed in the $c$-axis lattice constant. The changes below \TN{} are even more pronounced. In particular, the damping $\gamma$ of all these modes is abruptly reduced when entering the antiferromagnetically ordered phase. This overall decrease of $\gamma$ at \TN{} is relatively strong, for the modes at the lowest frequencies the damping changes by one order of magnitude.

\begin{figure}[tb]
\includegraphics[width = 1\columnwidth]{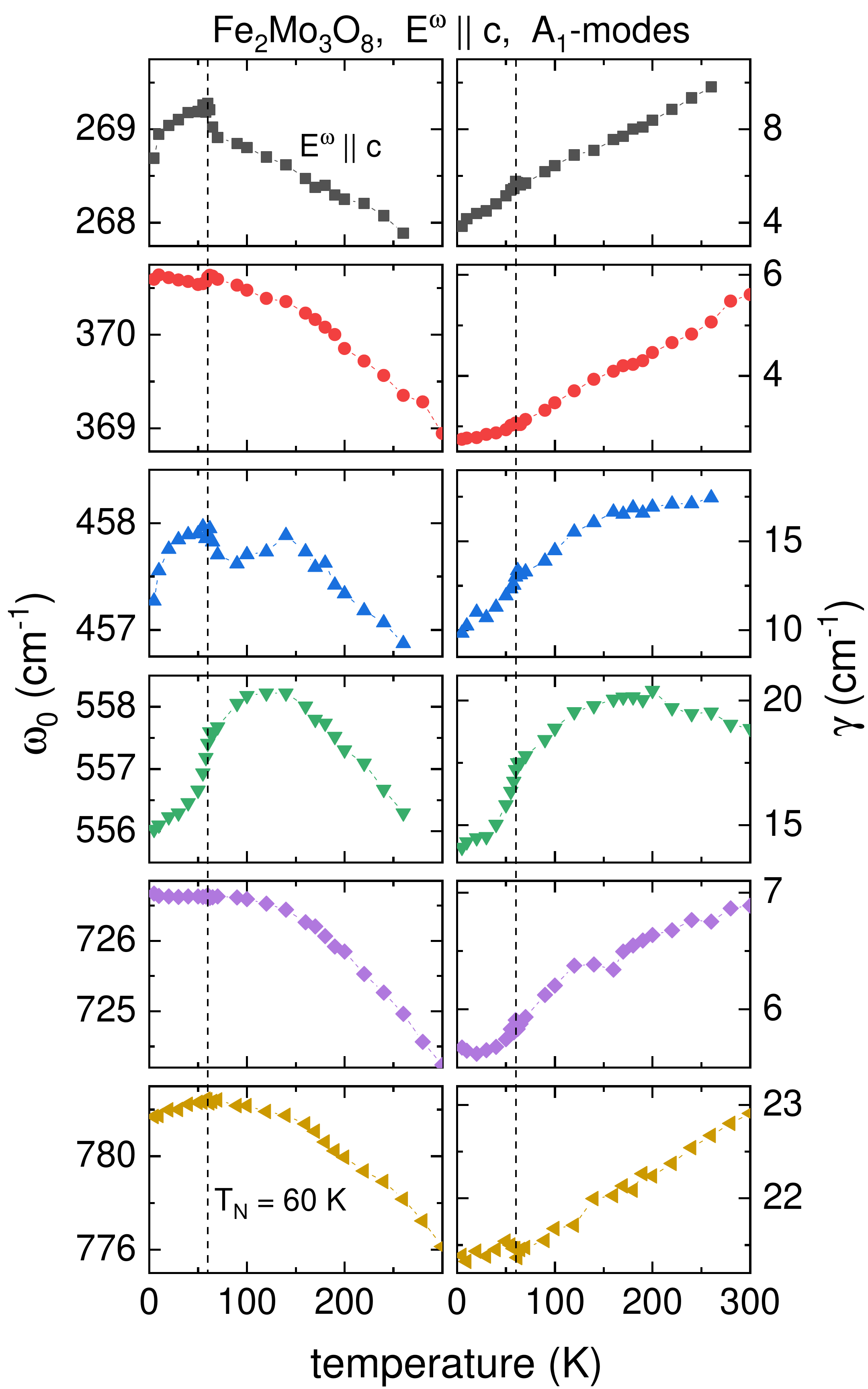}% Here is how to import EPS art
\caption{\label{fig:Fitparameter_Epara} Temperature dependence of phonon eigenfrequency $\omega_{0}$ and damping $\gamma$ for \Epara{}. The dashed vertical lines indicate the phase transition at $T_{\mathrm{N}}=60$\,K.}
\end{figure}

The temperature dependence of $\omega_0$ and $\gamma$ of the six strongest IR modes for polarization \Epara{} is shown in Fig.~\ref{fig:Fitparameter_Epara}. In this polarization direction, the phonons are less affected by the magnetic ordering. Above \TN{}, the eigenfrequencies of all modes monotonically increase with lowering the temperature, which is an expected behavior for anharmonic solids that stems from thermal expansion and phonon-phonon interactions~\cite{Cowley:1963,Cowley:1965,Klemens:1966,Menendez:1984}. Except the phonon mode located around 556\,cm$^{-1}$, the eigenfrequencies for \Epara{} experience either no or only slight changes at \TN{}. The 556\,cm$^{-1}$ phonon reveals a drop in $\omega_0$ below \TN{}, however, in this case the nearby \Eperp{} active mode, which can be seen due to polarization leakage, hampers the fitting. In contrast to the \Eperp{} active modes, the damping $\gamma$ shows an overall monotonically decreasing behavior with lowering temperature. Besides, $\gamma$ is not significantly affected by the magnetic ordering and shows a smooth behavior upon crossing \TN{}.

The hardening of the phonon modes upon cooling is also in line with the DFT+$U$+SO results obtained for the 1.7\,K and 275\,K crystal structures. Most of the modes are expected to harden~\cite{supplement}. For example, the frequencies of the $A_1$ modes should increase, on average, by 1\,cm$^{-1}$, which is indeed observed experimentally (Fig.~\ref{fig:Fitparameter_Epara}). A similar hardening trend is expected for the $E_1$ modes from the DFT+$U$+SO calculations, but most of them show a different behavior: some of them soften (135 and 290\,cm$^{-1}$), some remain temperature-independent (218 and 510\,cm$^{-1}$), or even behave non-monotonically (454 and 473\,cm$^{-1}$) between 300\,K and \TN{}. In fact, DFT+$U$+SO calculations for the antiferromagnetic spin configuration yield higher frequencies than in the ferromagnetic case~\cite{supplement}. Therefore, the onset of antiferromagnetic short-range order below 200\,K should generally harden the $E_1$ modes, which is not the case for at least half of them. However,  spin-phonon coupling \cite{Baltensperger:1968} can lead to such effects already in the paramagnetic phase of exchange coupled systems, as reported, for example, for ferrimagnetic FeCr$_2$S$_4$ \cite{Rudolf:2005}, or in frustrated spinel oxides \cite{Kant:2009}.

\begin{figure}[tb]
\includegraphics[width = 1\columnwidth]{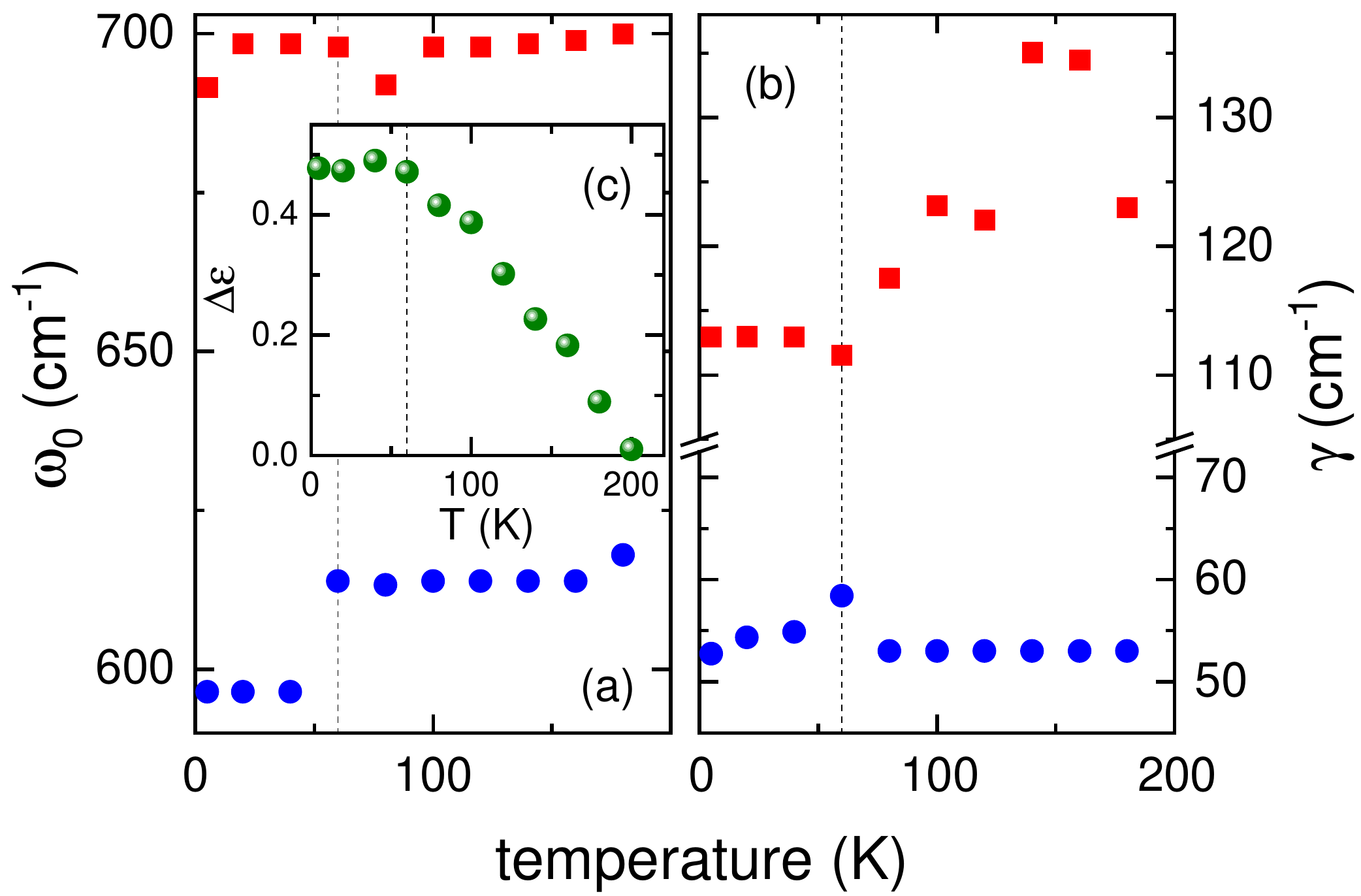}% Here is how to import EPS art
\caption{\label{fig:Fitparam_broad_feature} Temperature dependence of eigenfrequency $\omega_0$ (a), damping $\gamma$ (b) and oscillator strength $\Delta\epsilon$ (c) of the short-range order related absorption features observed for \Eperp{}.}
\end{figure}

We now discuss additional modes appearing in the spectra below 200\,K (group ii) and below \TN{} (group iii). The broad feature appearing below 200\,K and highlighted in Fig.~\ref{fig:FIR_R}c has been fitted with two Lorentzian oscillators. Fig.~\ref{fig:Fitparam_broad_feature} shows the temperature dependence of their $\omega_0$, $\gamma$, and the total oscillator strength $\Delta\epsilon$. No phonons of suitable symmetry are expected in this frequency range, and the spectral feature is indeed much broader than the typical phonon resonance. Its overall intensity follows the increase in the $c$-lattice parameter (Fig.~\ref{fig:nparam}b) and becomes saturated below \TN{}, suggesting short-range magnetic order as the origin of this feature and magnetic degrees of freedom closely involved. The eigenfrequency (Fig.~\ref{fig:Fitparam_broad_feature}a) of the upper mode is nearly constant between 5 and 200\,K within the experimental accuracy. For the lower-lying mode, $\omega_0$ exhibits a sudden drop when entering the antiferromagnetially ordered phase. The damping (Fig.\ref{fig:Fitparam_broad_feature}b) of the lower mode is constant within the error bar. For the excitation with higher frequency, $\gamma$ increases with increasing temperature above \TN{}.

Finally, several additional modes are observed below \TN{} only. Their frequencies are listed in the lower part of Table~\ref{tab:TAB_01}. There are three additional modes for \Eperp{} at 270, 426, and 468\,cm$^{-1}$ and one additional mode at 230\,cm$^{-1}$ for \Epara{} (arrows in Fig. \ref{fig:FIR_R}(a) and red dashed lines in Fig.~\ref{fig:FIR_sigma}). The eigenfrequencies and occurrence of these modes below \TN{} are in agreement with the spectra reported in Ref.~\cite{Stanislavchuk:2019}. Considering that the calculated eigenfrequencies of 286\,cm$^{-1}$ and 473\,cm$^{-1}$ have no direct experimental counterpart, the additional modes at 270\,cm$^{-1}$ and 468\,cm$^{-1}$ could correspond to two of the three missing $E_1$ modes of the hexagonal \fmo{} structure. However, the two other modes at 426\,cm$^{-1}$ and 230\,cm$^{-1}$ lie far away from any of the calculated frequencies. Since these modes appear at low temperatures only, it is tempting to assign them to a symmetry lowering below \TN{} as suggested in \cite{Stanislavchuk:2019}, but our diffraction data exclude this scenario. A third possibility is that these additional excitations are of mixed nature, because excitations of electronic origin are also possible in \fmo{}, as we discuss below.

\subsection{\label{sec:ddTrans}\ddt{} in the mid-infrared}

\begin{figure}[t]
\includegraphics[width = 1\columnwidth]{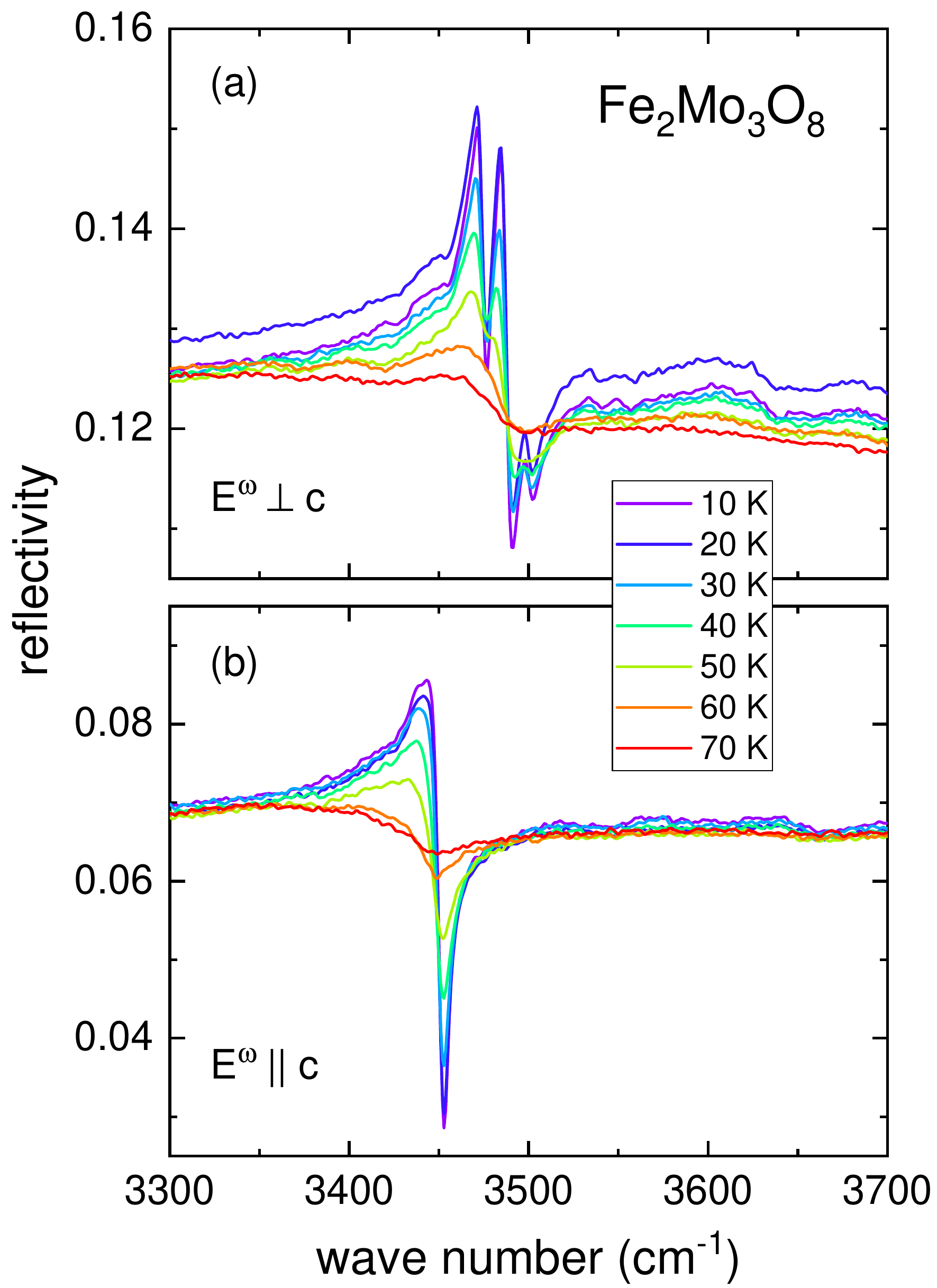}% Here is how to import EPS art
\caption{\label{fig:MIR_reflectivity} MIR reflectivity spectra in the temperature range between 10 and 70\,K for $E^{\omega} \perp c$ (a) and $E^{\omega} \parallel$ c (b) covering the frequency range of the \ddt{} of tetrahedral Fe$^{2+}$.}
\end{figure}

We complete our spectroscopic characterization by analyzing MIR reflectivity spectra for \Eperp{} and \Epara{}. The excitations observed between 3300 and 3700\,cm$^{-1}$ (Fig.~\ref{fig:MIR_reflectivity}) can be ascribed to the \ddt{} of the tetrahedral Fe$^{2+}$ ions \cite{Stanislavchuk:2019}, which are typically found in this frequency range~\cite{Low:1960a,Feiner:1982,Mualin:2002,Ohgushi:2005,Fedorov:2006,Laurita:2015,Evans:2017}. The excitation energy yields $\Delta_t\simeq 3500$\,cm$^{-1}$ (0.43\,eV), which is somewhat larger than found from DFT (0.33\,eV), probably due to the correlation and multiplet effects neglected in the calculation. 

For both polarization directions the \ddt{} show up as sharp features in the reflectivity at low temperatures. With increasing temperature the \ddt{} become gradually suppressed and considerably broadened when approaching \TN{}. Above \TN{}, the \ddt{} remain visible as weak and broad features in the reflectivity spectra. Thus, the temperature dependence of the linewidth of the \ddt{} strongly resembles that of the damping $\gamma$ for the phonon modes active for \Eperp{}. 

\begin{figure}[tb]
\includegraphics[width = 1\columnwidth]{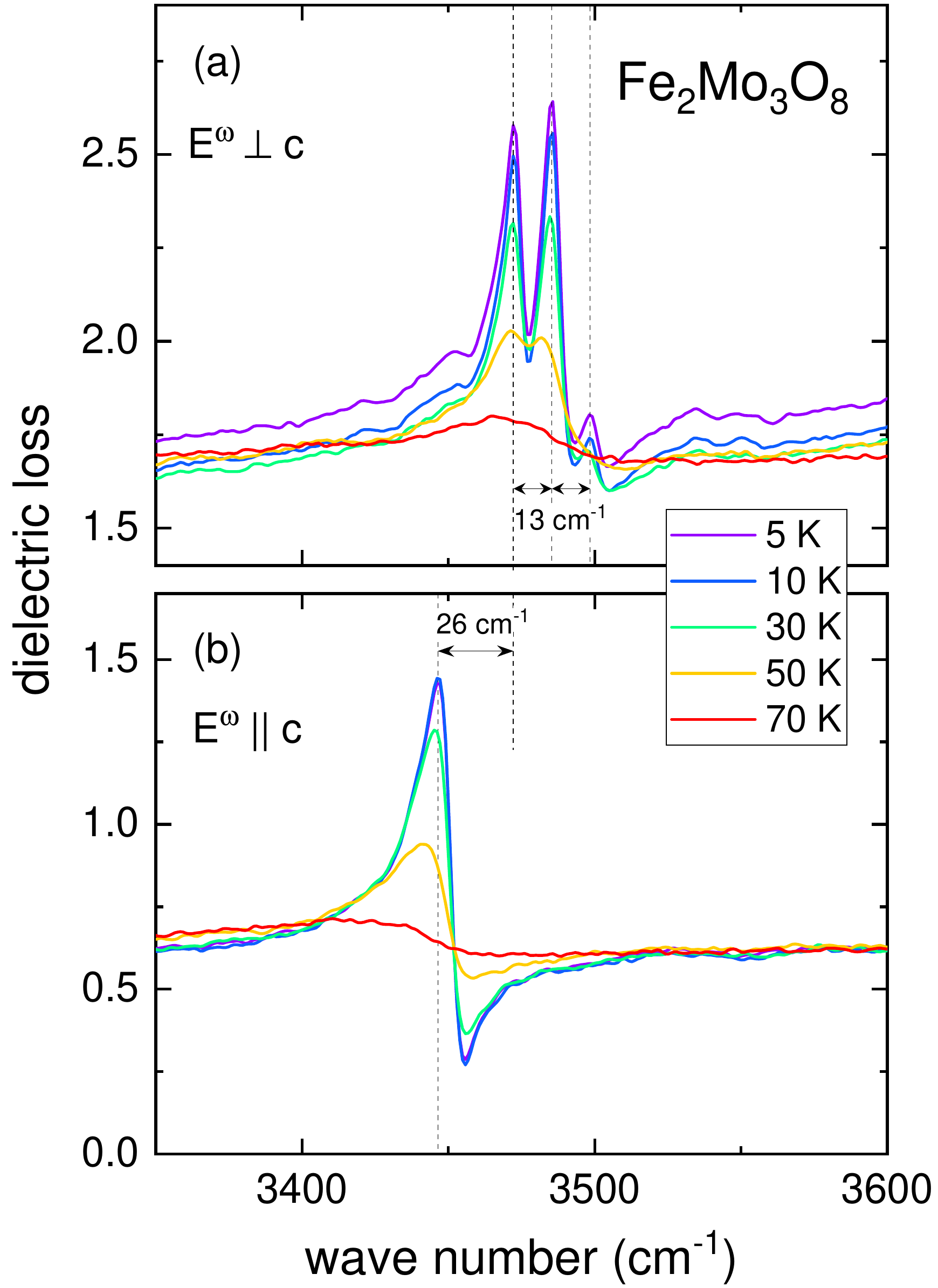}% Here is how to import EPS art
\caption{\label{fig:MIR_epsilon} Dielectric loss $\epsilon_2$ for \Eperp{} (a) and \Epara{} (b) in the frequency range of the \ddt{} of tetrahedral Fe$^{2+}$ for selected temperatures. Dashed lines indicate the energies of the $d$-$d$ excitations.}
\end{figure}

The \ddt{} show a clear selection rule. For \Eperp{} the spectrum is dominated by three excitations located at 3472, 3486 and 3498\,cm$^{-1}$, while for \Epara{} one excitation can be observed at 3447\,cm$^{-1}$. The excitation energies have been determined from the maxima of the dielectric loss $\epsilon_2$, which is shown in Fig. \ref{fig:MIR_epsilon} for \Eperp{} (a) and \Epara{} (b) for selected temperatures between 10 and 70\,K.

The energies of the mid-infrared excitations observed in our spectra and in Ref.~\cite{Stanislavchuk:2019} are summarized in Table~\ref{tab:dd-Trans}. The most intense excitations observed in MIR are in agreement with the data reported in Ref.~\cite{Stanislavchuk:2019}, but the additional sample-dependent broad excitation features in the reflectivity spectrum of Ref.~\cite{Stanislavchuk:2019} for \Eperp{} around 3000\,cm$^{-1}$ are absent in our spectra.

We want to point out that the distances of 13\,cm$^{-1}$  between the observed three excitations for \Eperp{} and the fact that the excitation observed for \Epara{} is separated from them by twice this value (26\,cm$^{-1}$) is in remarkable agreement with the level spacings of the excitation features in the frequency region 830-860~cm$^{-1}$ in Fig.~\ref{fig:FIR_sigma}. The presence of several \ddt{} stems from additional splittings introduced by the SO coupling into the level scheme shown in Fig.~\ref{fig:orbitals}. These splittings depend on the interplay of the SO coupling $\lambda$ and secondary crystal-field splitting $\delta_t$. With equally spaced \ddt{}, our data are reminiscent of the $\delta_t=0$ scenario~\cite{Low:1960,Slack:1966}, where adjacent levels of the ground multiplet should be separated by $6\lambda^2/\Delta_t$. Using the experimental separation of 13\,cm$^{-1}$ from Table~\ref{tab:dd-Trans}, we estimate the spin-orbit coupling constant $\lambda\simeq 87$\,cm$^{-1}$, which is on par with other compounds containing Fe$^{2+}$~\cite{Slack:1966,Laurita:2015}. Moreover, the direct observation of a low-energy excitation with an onset at 12\,cm$^{-1}$~\cite{Csizi:2020} in slightly  Zn-doped \fmo{} also suggests that this energy scale is generic for the Fe1 ground multiplet states. 

Interestingly, a point-charge model calculation of the level splittings in \fmo\,by Varret and coworkers \cite{Varret:1972} reveals a similar energy scale for the first excited state, although the authors primarily sought to explain the temperature variation of the observed M\"ossbauer data. In their model a two-fold degeneracy of the Fe1 ground state remains in the presence of spin-orbit coupling and it is lifted only by an internal exchange field present in the magnetically ordered state. However, the authors state that their fourth order trigonal crystal-field contribution is comparable to the cubic term $\Delta_t$, which is rather unusual and differs from our DFT results (Sec. IIIB). Clearly, it remains a future task beyond the scope of this study to calculate a detailed level scheme which consistently describes the optical excitations and the previously reported M\"ossbauer data.

\begin{table}[t]
\caption{\label{tab:dd-Trans}
Energies of the observed mid-infrared excitations at 5\,K in comparison to the observed Raman (R) and infrared-active (IR) modes reported in Ref. \cite{Stanislavchuk:2019}. The values are given in cm$^{-1}$. }
\begin{ruledtabular}
\begin{tabular}{ccc}
 This work & Ref. \cite{Stanislavchuk:2019} & Polarization\\
\hline
     & 3440 (R)    &  \\
3446 & 3448 (IR,R) & $E^{\omega} \parallel c$ \\
3472 & 3467 (IR,R) & $E^{\omega} \perp c$ \\
3485 & 3481 (IR,R) & $E^{\omega} \perp c$ \\
3498 & 3494 (IR,R) & $E^{\omega} \perp c$ \\
\end{tabular}
\end{ruledtabular}
\end{table}

\section{Discussion and Summary}
Several scenarios have been proposed for \fmo{} in the recent literature. In Ref.~\onlinecite{Stanislavchuk:2019}, symmetry lowering was inferred from the additional, presumably phonon modes appearing at low temperatures, whereas the authors of Ref.~\onlinecite{Solovyev:2019} speculated on the possibility of an orbital ordering, which should also lead to a symmetry lowering. Alternatively, they put forward charge separation between the octahedral and tetrahedral sites as a microscopic mechanism that does not require symmetry lowering~\cite{Solovyev:2019}. Our low-temperature structural data exclude all these possibilities. No signs of symmetry lowering have been observed down to 1.7\,K. Moreover, the local environment of both Fe sites does not change with temperature and remains typical for Fe$^{2+}$. The absence of charge separation is confirmed by the bond-valence-sum (BVS) analysis of interatomic distances~\cite{Brown:1985} that serves as a sensitive probe of Fe valence in mixed-valence oxides~\cite{Senn:2012,Ovsyannikov:2016}. Using experimental structural data at 1.7\,K, we estimate a BVS of 2.01 at the tetrahedrally coordinated site and 2.15 at the octahedrally coordinated site, which clearly rules out the Fe$^{1+}$--Fe$^{3+}$ scenario advocated by Ref.~\onlinecite{Solovyev:2019}. 

Our data also show that orbital degrees of freedom of the octahedrally coordinated Fe$^{2+}$ site are quenched by the trigonal distortion. This quenching does not occur for the tetrahedrally coordinated site. In our scenario spin-orbit coupling lifts the degeneracy and creates a significant orbital moment that contributes to the ordered magnetic moment below \TN{}. With the experimental value of 4.61(2)\,$\mu_\mathrm{B}$ at 1.7\,K, the ordered moment clearly exceeds its spin-only value. It confirms the formation of the orbital moment on Fe1 and, consequently, suggests that spin-orbit coupling is the primary mechanism to lift the orbital degeneracy of the tetrahedrally coordinated Fe$^{2+}$ site in \fmo{}.

The scenario of a large orbital moment and absent Jahn-Teller distortion was previously reported for isolated Fe$^{2+}$ centers in ZnS and other semiconductors~\cite{Slack:1967}. In periodic systems like Fe$^{2+}$ thiospinels Jahn-Teller distortions can occur but only at very low temperatures~\cite{Tsurkan:2010,Deisenhofer:2019}, and show a negligibly small orbital contribution to the magnetic moment~\cite{Bertinshaw:2014}. The absence of a Jahn-Teller distortion in \fmo{} may be related to the large separation between the Fe$^{2+}$ ions and the simultaneous presence of octahedrally coordinated ions, which are not prone to any distortion.

The anomalous behavior of the $E_1$ phonon modes, including the non-monotonic evolution of their frequencies and abrupt increase in the phonon lifetime at \TN{}, may be related to orbital degrees of freedom too. The fact that these features are only observed in the $E_1$ channel suggests their relation to the electronic levels of the $e_1$ symmetry, where orbital degeneracy occurs. Indeed, orbital fluctuations can occur through the fluctuating direction of the orbital moment. It should be parallel to the spin moment of Fe$^{2+}$, but otherwise is not constrained. In the absence of magnetic order, the orbital moment of Fe$^{2+}$ can point along $c$ or along $-c$, which are two different and distinguishable directions in the polar crystal structure of \fmo{}. Internal fields that set in below \TN{} fix the direction of the spin moment on a given Fe site and, therefore, quench orbital fluctuations. This quenching may affect the lifetime of the $E_1$ phonons.

Concerning the number of phonon modes, we identified eight out of the nine expected $A_1$-modes showing an excellent agreement with our DFT+$U$+SO calculation and nine out of twelve $E_1$ modes with a less satisfactory agreement. The additional weak modes in the range of 830-860\,cm$^{-1}$ and the broad absorption feature at 600-700\,cm$^{-1}$ undergo changes at around 200~K, indicating their relation to the minimum in the temperature dependence of the lattice parameter $c$ and the associated onset of short-range magnetic order. None of these modes can be interpreted as pure phonons. They probably have a combined vibrational and electronic origin.

The additional modes for \Eperp{} and \Epara{} observed below $T_N$ may not be pure phonons either, as magnetic degrees of freedom allow for combined phonon and magnon excitations. Moreover, it may be difficult to track experimentally, whether these modes are present in the magnetically ordered state only, or exist at all temperatures and simply become visible around \TN{}, because their lifetime increased, similar to the $E_1$ phonons (Fig.~\ref{fig:Fitparameter_Eperp}). In this case, a combination of electronic and phonon excitations becomes another plausible scenario. The SO coupling on the tetrahedral Fe$^{2+}$ site indeed creates low-energy excitations that are optically allowed~\cite{Slack:1967} and couple to phonons, sometimes in a very intricate way~\cite{Slack:1969}. 
%An unambiguous assignment of these coupled modes appears to be difficult even for single Fe$^{2+}$ ions~\cite{Slack:1969,Ham:1971,Vogel:1980}, let alone for the concentrated Fe system of \fmo{} with additional magnetic degrees of freedom appearing below 200\,K, where short-range magnetic order sets in. 
%Our finding of equally spaced $d$-$d$ excitations around 3500\,cm$^{-1}$ and the excitations around 830-860\,cm$^{-1}$ indicate an unusual electronic level scheme of the tetrahedral Fe$^{2+}$ sites, which might originate from vibronic couplings.

In summary, our crystallographic study of the \fmo{} multiferroic confirmed its robust hexagonal symmetry down to low temperatures. Orbital degrees of freedom are quenched on the octahedrally coordinated Fe$^{2+}$ site but remain active on the tetrahedrally coordinated site, where spin-orbit coupling generates a sizable orbital moment. The anomalous temperature dependence of the $E_1$ phonon modes with their largely increased lifetime in the magnetically ordered state may also indicate the importance of orbital degrees of freedom, possibly via constraining the orbital moment direction in the antiferromagnetic state. The electronic excitations associated with the Fe$^{2+}$ ions on the tetrahedral sites suggest a splitting of the ground and first excited state by 13\,cm$^{-1}$, in agreement with the direct observation of a low-energy mode, which was reported recently for Zn-doped \fmo{} and interpreted in terms of a vibronic excitation \cite{Csizi:2020}.

\begin{acknowledgements}
We thank ALBA and PSI for granting the beamtime for this project and acknowledge Aleksandr Missiul (ALBA) and Denis Sheptyakov (PSI) for their help with the data collection. AAT and NK were supported by the Federal Ministry for Education and Research through the Sofja Kovalevskaya Award of Alexander von Humboldt Foundation. This research was partly funded by Deutsche Forschungsgemeinschaft DFG via the Transregional Collaborative Research Center TRR 80 “From Electronic correlations to functionality” (Augsburg, Munich, Stuttgart).
\end{acknowledgements}

\providecommand{\noopsort}[1]{}\providecommand{\singleletter}[1]{#1}%

\end{document}